\documentclass{aa}
\usepackage{graphicx}
\usepackage{natbib}
\usepackage{txfonts}

\def\hess{{\em HESS}}
\def\ls{LS~5039}
\def\lsi{LSI~+61$\degr$303}
\def\psrb{PSR~B1259-63}
\def\xte{XTE~J1550-564}
\def\spose#1{\hbox to 0pt{#1\hss}}
\def\la{\mathrel{\spose{\lower 3pt\hbox{$\mathchar"218$}}
     \raise 2.0pt\hbox{$\mathchar"13C$}}}
\def\ga{\mathrel{\spose{\lower 3pt\hbox{$\mathchar"218$}}
     \raise 2.0pt\hbox{$\mathchar"13E$}}}

\begin{document}
\title{Gamma-ray absorption in massive X-ray binaries}
\author{Guillaume Dubus
	\inst{1,2}
	}

\institute{Laboratoire Leprince-Ringuet, UMR 7638 CNRS, Ecole Polytechnique, 91128 Palaiseau, France; \email{dubus@in2p3.fr}
\and
Institut d'Astrophysique de Paris, UMR 7095 CNRS, Universit\'e Pierre \& Marie Curie, 98 bis boulevard Arago, 75014 Paris, France
}

%\date{Draft \today}
\abstract
{Gamma-ray emission in the TeV ($10^{12}$~eV) range has been detected by \hess\ from two X-ray binaries: \psrb\ and \ls. In both, the early-type star provides large numbers of target photons for pair-production with TeV $\gamma$-rays. This results in a modulation of the $\gamma$-ray flux as the relative positions of the $\gamma$-ray source and companion star change with orbital phase for the observer.
The extent to which this variable absorption can provide useful diagnostics for the location and nature of $\gamma$-ray emission is examined.
The absorption spectrum and transmitted flux are calculated by integrating the $\gamma\gamma$ cross-section along the line-of-sight, taking into account the orbit, the spectrum and the finite size of the companion star in \ls, \psrb\ and \lsi, a system similar to \ls\ but still undetected at TeV energies.
In \ls, emission close to a black hole or a neutron star primary is considered. In both cases, the transmitted flux $>$250~GeV drops by an order-of-magnitude near periastron ($\phi$=0). A black hole yields a clear spectral signature in the average spectrum at $\approx 400$~GeV. A neutron star yields more variability, with the spectral feature moving from 200~GeV ($\phi$=0.1) to 3~TeV ($\phi$=0.7). Only 20\% of the flux is absorbed at $\phi$=0.7, allowing for an almost direct view of the intrinsic spectrum. Low variability will require emission on large scales, more than 0.7~AU away to have $<$50\% absorption in a jet. In \lsi, significant absorption (up to 90\% of the 100~GeV flux) is predicted only slightly before periastron, accompanied by a spectral hardening above 1~TeV. In \psrb, although 40\% of the flux is absorbed before periastron, the large variability seen by \hess\ is due to the $\gamma$-ray emission process.
 The predictions made here are essential to distinguish variability in the emission of $\gamma$-rays from that due to absorption. A modulation would provide a novel way to constrain the $\gamma$-ray source. Its absence would imply that $\gamma$-ray emission occurs on large scales.
}
\maketitle

\section{Introduction}
Massive X-ray binaries, composed of a neutron star or black hole in orbit around an early-type star, have long been recognized as possible $\gamma$-ray emitters \citep{weekes}.  Recently, the \hess\ collaboration has reported very high energy (VHE) $\gamma$-ray emission in the TeV (10$^{12}$ eV) range from two such systems. The first, \object{\psrb} (=\object{SS 2883}) is a 47.7~ms radio pulsar in a 3.4~yr orbit around a B2Ve star. Variable $\gamma$-ray emission was detected by \hess\ during periastron passage \citep{psrb}. Here, the relativistic pulsar wind is confined by the stellar wind from the massive companion. Particles at the termination shock produce the non-thermal emission \citep{tavani}. The second, \object{\ls} (=\object{RX J1826.2-1450}), is a tight 4-day binary composed of an O6V star and an unknown compact object.  Resolved radio emission from the source is thought to arise from a weakly relativistic jet with $v\approx0.3c$; \ls\ has also been associated with the EGRET source 3EG 1824-1514 \citep{paredes}. This suggests \ls\ might be a scaled-down analogue of extragalactic blazars. The VHE emission is hard (power law photon index $\approx -2$) and continuously detected over a timespan of 3 months in 2004 \citep{science}.

The large stellar luminosity ($2~10^5~L_\odot$ in \ls) provides an abundant source of seed photons for inverse Compton scattering with high energy particles. At the same time, the dense radiation field can absorb very high energy $\gamma$-ray emission through pair production $\gamma\gamma\rightarrow e^+e^-$ \citep{gs}. With stellar effective temperatures in the 20,000-40,000~K range, the typical target photon energy is a few eV which is in the energy range for pair production with TeV photons. The cross-section maximum occurs close to the threshold energy and has an amplitude $\sigma_{\gamma\gamma}\approx \sigma_T/5$ where $\sigma_T$ is the Thomson cross-section. The star luminosity is very high ($10^{38}-10^{39}$ erg~s$^{-1}$) for radii $\sim 10$ R$_\odot$. At a distance $d\approx$ 0.1~AU corresponding to periastron in \ls\ (see \S3), the stellar radiation density in 3.5~eV photons is $n_\star\approx 10^{14}~\mathrm{photons~cm}^{-3}$. A rough estimate of the $\gamma$-ray opacity is $\tau_{\gamma\gamma}\approx \sigma_{\gamma\gamma}n_\star d\approx 20$.  The detection of VHE photons from \ls\ is therefore intriguing. 

The $\gamma\gamma$ opacity depends strongly on the geometry. The opacity can be greatly reduced compared to the simple estimate above because line-of-sight $\gamma$-rays travel in an anisotropic radiation field. Diffusion in the wind is too small to significantly isotropize radiation \citep{dermer}. The absorption will vary depending upon the relative location of the source of $\gamma$-rays, the companion star and the observer. In binary systems, the orbit and companion star type can be constrained by optical observations. If $\gamma$-ray emission is isotropic and close to the compact object, absorption will be periodically modulated in a predictable way \citep{protheroe,moskalenko}. Contrast this situation with that in blazars, where the sources of radiation (accretion disc, broad line region) that can absorb $\gamma$-rays close to their production site are poorly known. A generic overview of $\gamma\gamma$ absorption in binaries is presented in \S2.

The effects of $\gamma$-ray absorption are predicted in \S3-4 for \ls, \psrb\ and \object{\lsi} (= \object{V615~Cas}), a system very similar to \ls. Comparisons with existing or future observations (will) provide useful diagnostics of the location of $\gamma$-ray emission and its nature. \citet{dermer} present calculations for \ls, focusing on emission from a black hole jet, and assuming $e=0$ and a point source approximation for the star. Here, expected differences in absorption lightcurves and spectra for $\gamma$-ray emission close to a black hole or a neutron star primary are examined for \ls\ and \lsi\ using the measured orbits and taking into account the finite star size. Expectations if emission arises from a jet are considered in \S5. In this work, the $e^+e^-$ pairs created by absorption are assumed to escape the system or radiate at much lower frequencies. Possible caveats when the pairs radiate back in $\gamma$-rays and initiate an electromagnetic cascade are discussed in \S6.

\section{VHE $\gamma$-ray opacity around massive stars}
\begin{figure*}
\resizebox{17cm}{!}{\includegraphics{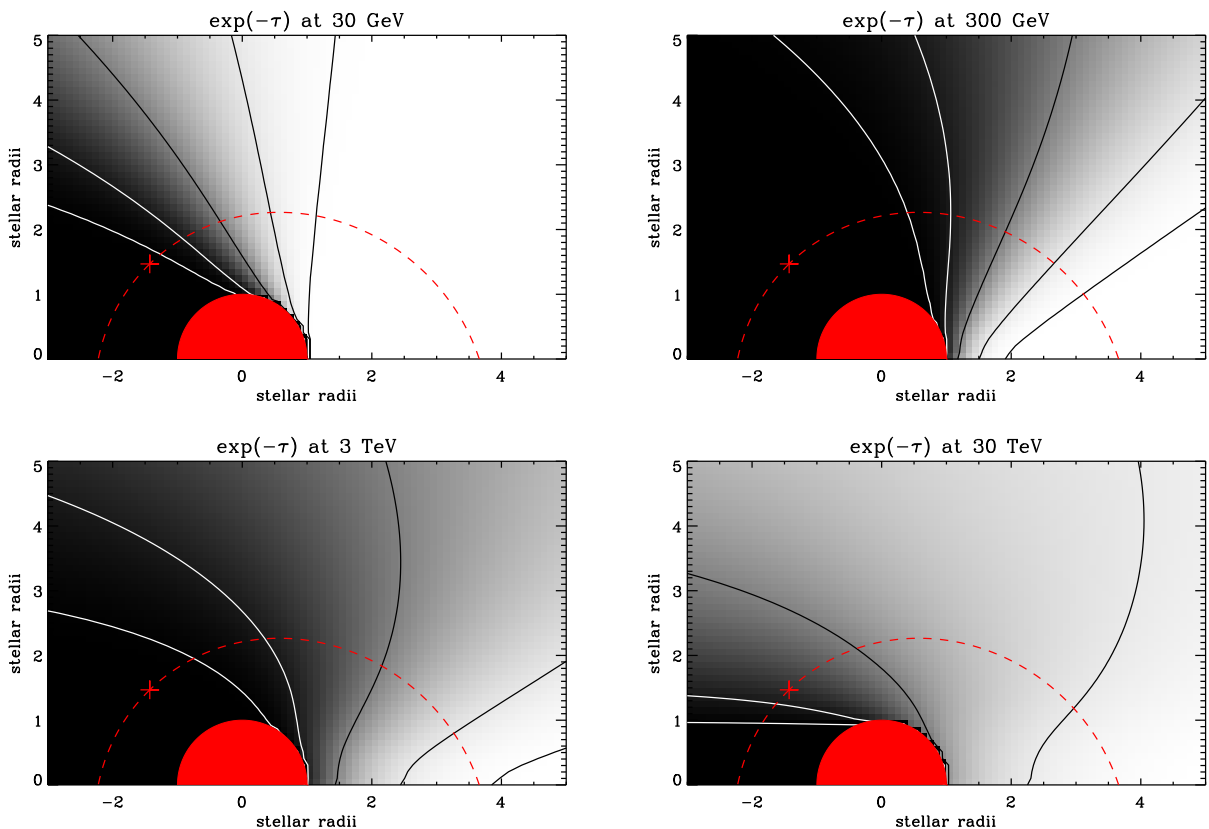}} 
\caption{Gamma-ray absorption map around a massive star. The grey-scale at each location corresponds to the integrated $\gamma\gamma$ absorption seen by a VHE photon emitted at this location and travelling in the plane of this figure towards an observer located at infinity to the right. The plotted value is $\exp(-\tau_{\gamma\gamma})$, on a linear scale with black corresponding to high absorption and white to low absorption. Absorption is shown for VHE photons with energies of 30~GeV, 300~GeV, 3~TeV and 30~TeV. Full lines are contours corresponding to 99\%, 90\% (white), 50\%, 10\% and 1\% (black) of the emitted flux being absorbed. The star is located at the origin, has a 10 $R_\odot$ radius and an effective temperature of 40,000~K. The orbit of the binary system \ls\ is (partially) shown as a dotted line for illustrative purposes. Here, the binary would be seen edge-on ($i$=90\degr) by the observer at right. Gamma-rays emitted close to the compact object towards this observer are more or less absorbed according to orbital phase, with maximum absorption occurring after periastron (marked by a cross). The resulting flux modulation would be close to that shown in Fig.~\ref{ls5039lc} for the case of $i$=60\degr.}
\label{maps}
\end{figure*}
The absorption of a $\gamma$-ray of energy $E$  on a star photon of energy $\epsilon$ occurs above an energy threshold given by \citep{gs}
\begin{equation}
\epsilon E\ge \frac{2 m_e^2 c^4}{(1-\mathbf{e}_\gamma \mathbf{e}_\star)}
\label{eq:threshold}
\end{equation}
where $\mathbf{e}_\gamma$ is a unit vector along the direction of propagation of the $\gamma$-ray; $\mathbf{e}_\star$ is the unit vector along the direction of propagation of the star photon (see Appendix, Fig.~\ref{geometry}). Absorption can only occur when the target (stellar) photons are at an angle to the $\gamma$-ray and the minimum $m_e^2 c^4$ occurs for head-on collisions. 

The cross-section maximum occurs close to the threshold, a useful feature for quick estimates but one that errs as the angular dependence has a significant effect. For instance, $\gamma$-rays propagating directly away from the source of target photons are not absorbed (the energy threshold given by Eq.~\ref{eq:threshold} is infinite). The decrease in the radiation field density when integrating over the photon path also reduces the total opacity. The opacity seen by a $\gamma$-ray of given energy will therefore depend strongly on its emission location with respect to the star in the binary. 

The opacity depends also on the spectrum of target photons. High energy $\gamma$-rays are able to interact with the stellar Rayleigh-Jeans tail while lower energy $\gamma$-rays interact with the exponentially decreasing Wien part. At periastron, the compact object in \ls\ is only $\approx$ 1 $R_\star$ away from the surface of the star (Fig.~\ref{maps}). The finite size of the star can change results from a point source approximation \citep[e.g.][]{moskalenko}, notably by giving a non-zero opacity for a radially outgoing $\gamma$-ray. All these effects can be accurately computed (Appendix).

\subsection{Angular dependence}
A map of absorption around a star is shown in Fig.~\ref{maps} for various $\gamma$-ray energies. The star has an effective temperature $T_\star= 40,000$~K and radius $R_\star=10~{\rm R}_\odot$, appropriate for the O6.5V star in \ls\ (\S3). The grey scale represents the amount of flux absorption $\exp(-\tau_{\gamma\gamma})$ depending upon location of  $\gamma$-ray emission. The $\gamma$-rays are emitted in the plane of the figure, towards an observer situated at infinity to the right. Note that the solution has cylindrical symmetry about the line-of-sight to the star center. 

At low energies, the opacity is small until the $\gamma$-ray has an energy above the minimum threshold for head-on interaction with stellar photons $kT_\star$. The opacity then rises quickly as the target radiation density initially increases exponentially. At progressively larger $\gamma$-ray energies, the dominant effect in setting the opacity becomes the angular dependence (top panels).

At a given $\gamma$-ray energy, there is a cone pointing outwards close to the star for which $\tau_{\gamma\gamma}$ is reduced. The opening angle roughly corresponds to the threshold for interaction with stellar photons at the Wien cutoff. Keeping $\epsilon(1-\mathbf{e}_\gamma \mathbf{e}_\star)$ constant in Eq.~\ref{eq:threshold} and starting on the $x$-axis, $\gamma$-rays interact with an exponentially increasing density of stellar photons as the angle increases. However, because of the finite size of the star, the absorption can still be high in this cone if the $\gamma$-ray is emitted close to the star. For example, $>$50\% of the 3 TeV photons emitted outwards $<$0.5 $R_\star$ away from the star are still absorbed (bottom left panel of Fig.~\ref{maps}).

As the energy of the $\gamma$-ray further increases (bottom panels), the dominant effect in modifying the opacity is the decrease in radiation density since the interaction now occurs with Rayleigh-Jeans photons. The decrease in opacity is then roughly independent of location. The situation is slightly more complex on-axis as the angular dependence is always important there.

\subsection{Orbital variability}
The existence of a cone with low $\gamma$-ray opacity shows that VHE emission can occur close to the massive star without significant absorption. However, this configuration is variable as the binary aspect changes for the observer. Assuming isotropic emission close to the compact object, in a small region compared to the binary size, the resulting absorption lightcurve at a given energy depends on the orbital parameters (orbital period $P_{\rm orb}$, total mass $M$, eccentricity $e$, periastron longitude $\omega$ and inclination $i$, see Appendix). Knowledge of the orbital parameters obtained, for example, via pulsar timing or radial velocity studies, leads to constraints on the physics of VHE emission by comparing expected and observed lightcurves.

The case for which absorption is invariant of orbital phase is for a face-on, circular  ($i=e=0$). An inclined, circular orbit results in absorbed flux lightcurves which are symmetric with respect to the orbital phase of superior or inferior conjuction. The changing orbital separation in an eccentric orbit varies the absorption even with $i=0$\degr. Eccentric orbits typically result in $\gamma$-ray lightcurves with a flux dip at periastron.

\subsection{Absorbed spectrum}
The maps in Fig.~\ref{maps} show that the extinction depends on
location but also on energy. Pure $\gamma$-ray absorption produces an
absorption trough in the spectrum that, if measured, can provide
important diagnostics for emission processes \citep{dermer}. The
location of maximum absorption changes as the compact object revolves
around the massive star, providing further diagnostic power.  The
trough is quite wide, becoming obvious only for data covering a large
span in energies. For temperatures typical of massive stars, this span
is $\sim$ 30~GeV--30~TeV (\S3). The emitted $\gamma$-ray spectrum is
softened by absorption in data limited to the low energy end
($\la$300~GeV) while it is hardened at high energies
($\ga$300~GeV). In contrast, the spectral energy distribution of the
extragalactic background light results in a softening at TeV energies
of the emitted $\gamma$-ray spectrum from blazars
\citep{stecker}. Gamma-ray absorption by Galactic background radiation
is negligible for X-ray binaries.

\section{Gamma-ray absorption in \ls}
The origin of the VHE emission detected by \hess\ from \ls\ has yet to be established. An intriguing possibility is that the emission occurs in a relativistic jet as in blazars. Resolved radio emission in \ls, attributed to a mildly relativistic jet, constitutes a strong argument in favour of this possibility. Alternatively, the VHE emission could be a result of the interaction of a pulsar relativistic wind with the stellar wind from the massive star, as in \psrb\ \citep{maraschi,tavani}. The similarity in spectral energy distributions between \ls\ and \psrb\ argues in favour of this interpretation. 

Can the expected $\gamma$-ray absorption be used to investigate the origin of the $\gamma$-ray emission ? The star temperature and radius derived from optical spectroscopy are $T_\star\approx39,000$~K and $R_\star\approx9.3$~R$_\odot$ (\citealt{casares}, thereafter C05). The star mass is $\approx 23$ M$_\odot$. Radial velocity studies provide a good determination of $P_{\rm orb}\approx 3.906$~days, eccentricity $e\approx0.35$ and periastron angle $\omega\approx226$\degr. The orbital solution of C05, which includes data from \citet{mcswain} whilst adding better phase coverage, supersedes previous determinations.  By assuming that the secondary is pseudo-synchronized (in co-rotation at periastron), C05 find $i\approx25$\degr\ and a compact object mass $3.7\pm 1.3$~M$_\odot$. Whether the star has had time to pseudo-synchronize in this young, massive X-ray binary is not clear, and a neutron star cannot be firmly ruled out yet. Taking into account the mass function and the more conservative constraints on the system inclination $i$ derived by C05, two orbital solutions are adopted here to illustrate the possible extremes: one with $i=60$\degr, corresponding to a canonical 1.4~M$_\odot$ neutron star; and one with $i=20$\degr, implying a 4.5~M$_\odot$ black hole.

Hereafter, the $\gamma$-ray emission is assumed constant, isotropic and occuring in a small region (compared to the binary separation) close to the compact object. Pairs created by the absorption of $\gamma$-rays are further assumed to escape freely or to radiate their energy at frequencies below those of interest. These assumptions are examined in \S5-6. The aims are to investigate (1) how much absorption occurs and at what phases; (2) expected observational signatures; (3) whether a black hole or a neutron star primary can be distinguished.

\subsection{Orbital variation of absorption}

\begin{figure}
\resizebox{\hsize}{!}{\includegraphics{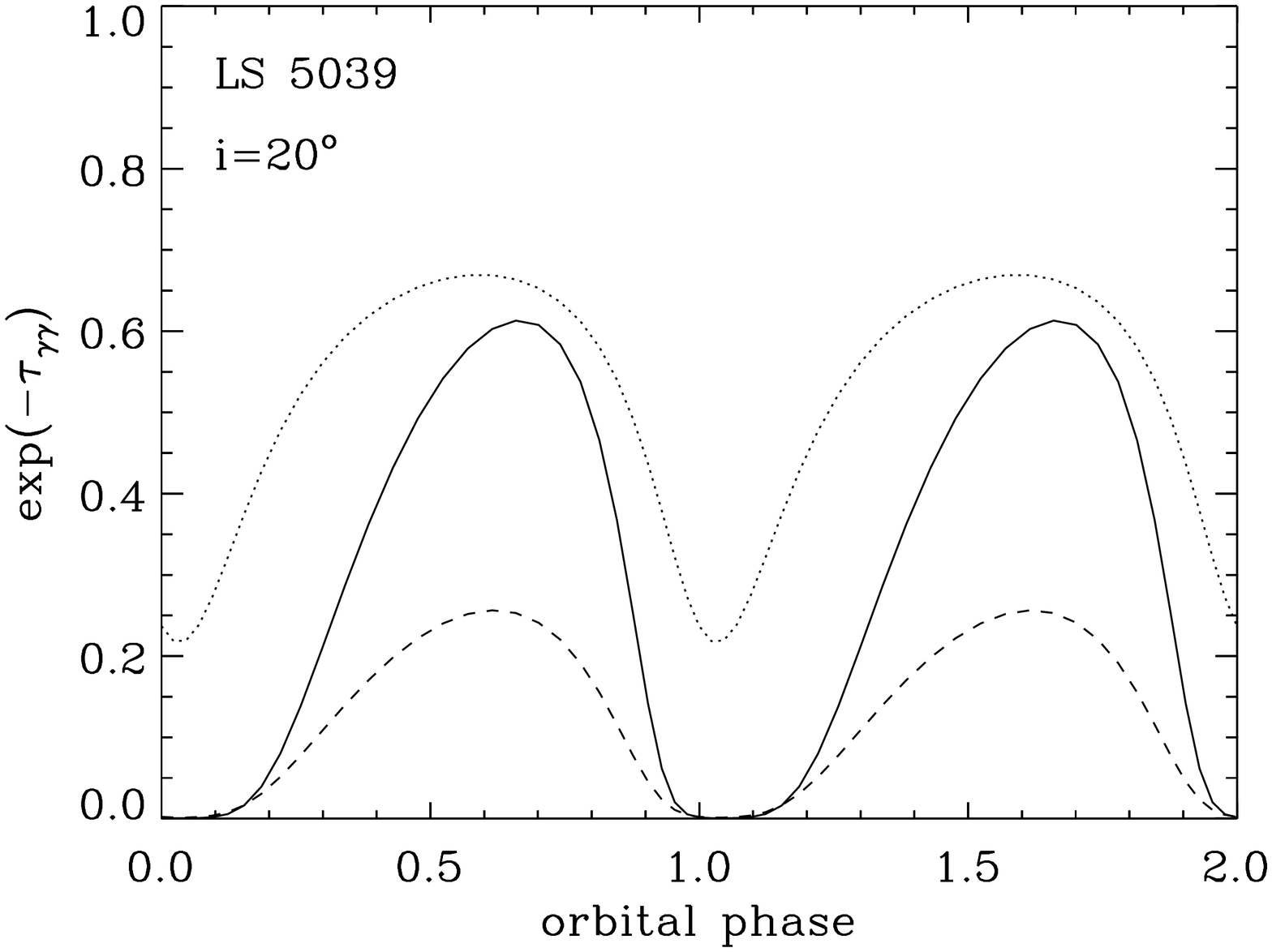}}
\resizebox{\hsize}{!}{\includegraphics{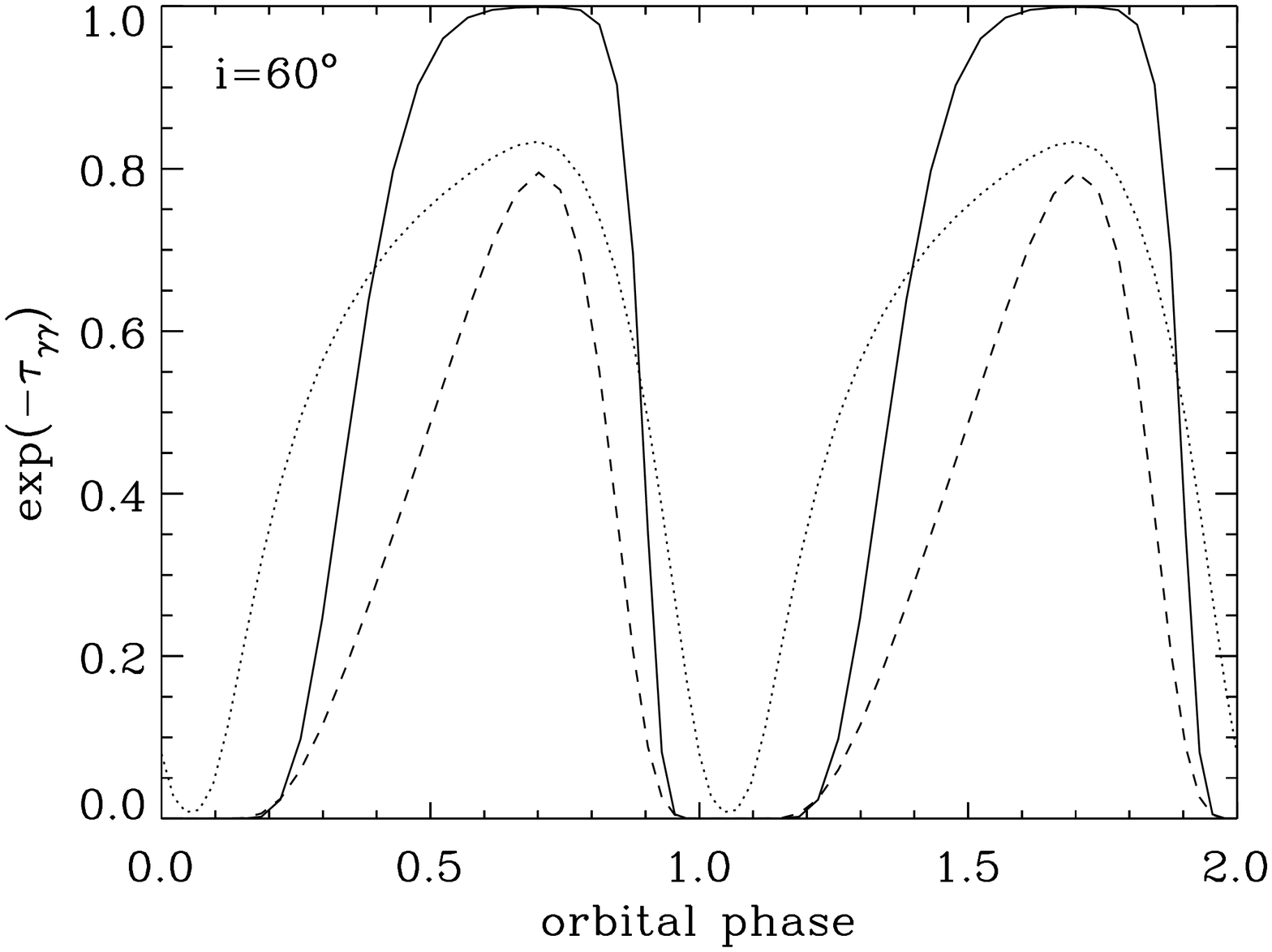}}
\caption{Modulation of the $\gamma$-ray transmission
$\exp(-\tau_{\gamma\gamma})$ with orbital phase in LS~5039 for two
values of the system inclination $i$. A canonical neutron star
corresponds to $i=60$\degr\ while $i=20$\degr\ implies a
$4.5$~M$_\odot$ black hole. Isotropic $\gamma$-ray emission is assumed to
occur at the location of the compact object. The solid line is for
100~GeV photons (H.E.S.S. threshold), dashed is for 1~TeV and dotted
for 10~TeV photons. Periastron passage is at $\phi=0$.}
\label{ls5039lc}
\end{figure}
\begin{figure}
\resizebox{\hsize}{!}{\includegraphics{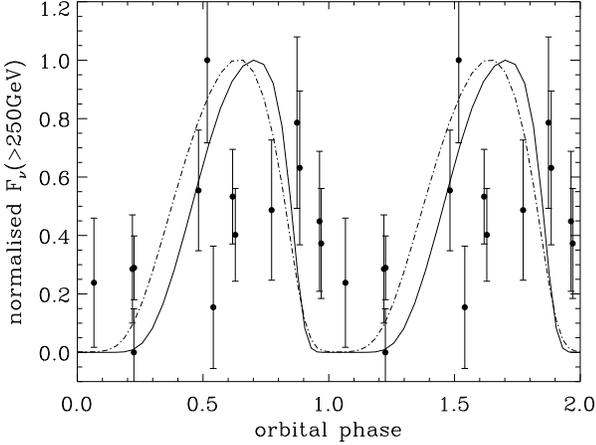}}
\caption{\hess\ observed lightcurve of \ls\ in 2004 (points with error bars, \citealt{science}), folded on the orbital period using the ephemeris of C05. The expected lightcurve caused by absorption is plotted as a solid line for $i=60$\degr\ and as a dashed line for $i=20$\degr. Here, the lightcurves are {\em integrated} above 250~GeV and normalized to their maximum (Fig.~\ref{ls5039lc} showed the absorption for given energy). The orbital phase of the peak and the duration of strong absorption ($>90$\%) are different between emission close to a black hole ($i=20$\degr) and close to a neutron star ($i=60$\degr).}
\label{ls5039int}
\end{figure}
\begin{figure}
\resizebox{\hsize}{!}{\includegraphics{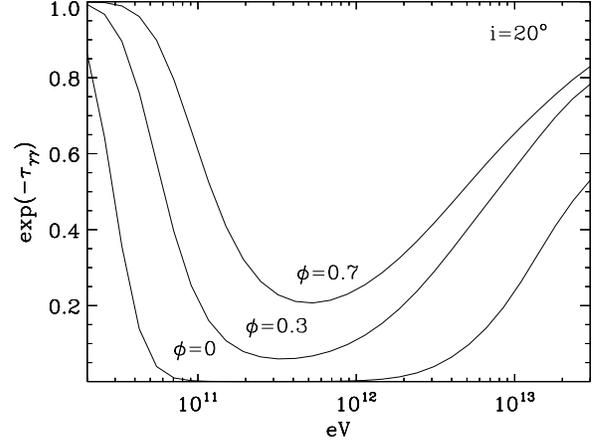}}
\resizebox{\hsize}{!}{\includegraphics{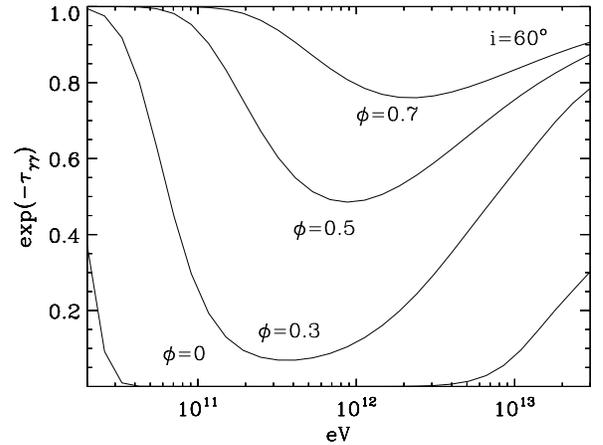}}
\caption{Spectrum of $\gamma$-ray absorption at selected orbital phases $\phi$ for two inclinations $i$ of \ls\ (see Fig.~\ref{ls5039lc}). The absorption spectrum varies little for $i=20$\degr. The line-like absorption feature may therefore be easier to detect for low inclinations, despite the stronger absorption near periastron for $i=60$\degr. In the latter case, although spectral variations will be easier to detect, the large differences in absorption spectrum can swamp out the line feature in a phase-averaged spectrum. An almost absorption-free view of the intrinsic spectrum is expected at flux maximum for $i=60$\degr.}
\label{ls_spec}
\end{figure}

The orbital variation of $\gamma$-ray absorption at various energies
is shown in Fig.~\ref{ls5039lc} for $i=20$\degr\ and $i=60$\degr. In
both cases, significant absorption in the 0.1-1~TeV band is predicted
at periastron. The opacity reaches $\tau_{\gamma\gamma}\approx 40$ at
phase $\phi\approx0.1$ for 200~GeV photons ($i=60$\degr). However, the
opacity is smaller than unity over a large fraction of the orbit. VHE
$\gamma$-rays escape because of the anisotropy of the target radiation
field. The minimum absorption (peak flux) occurs at $\phi\approx 0.6$
when the compact object is far from the star and close to the
observer. 

A non-detection of the periodic variation by \hess\ would suggest that
$\gamma$-ray emission occurs far from the companion star. Indeed, the
available \hess\ lightcurve from 2004 is statistically constant at the
15\% confidence level \citep{science}. A periodogram shows no peak
with more than 2$\sigma$ significance \citep{science}. However, only
14 nightly flux averages over 120 days are available. Future
observations will allow firmer conclusions to be drawn.

The amplitude and asymetry of the variation decrease with
inclination. However, these differences may be difficult to
distinguish when comparing to observations of lightcurves {\em
integrated} above the instrumental enegy threshold. The absorbed
fluxes for the two inclinations are plotted in Fig.~\ref{ls5039int}
along with the \hess\ 2004 lightcurve from \citet{science}. The
intrinsic spectrum was assumed to be a power law $dN_\gamma\propto
E^{-2.4}dE$, which was integrated after absorption above the same
threshold as the \hess\ data (250~GeV) to yield the expected flux
variations. The average expected photon index is consistent in both
cases with the \hess\ measurement (-2.12$\pm$0.15,
\citealt{science}). For an inclination of 20\degr\ (black hole), the
average fraction of the intrinsic flux that is absorbed above 250~GeV
is 89\%. For $i=60$\degr\ (neutron star), the fraction is 64\%. In
Fig.~\ref{ls5039int}, the observed and expected lightcurves are
renormalised to their maximum. The length of the absorption `eclipse'
is longer and the peak flux occurs later in phase for a high
inclination.

\subsection{Absorption spectrum}

The average absorption is greater for low inclinations as the angle between the star, emission location and observer ($\psi$, Fig.~\ref{geometry}) stays close to 90\degr. Therefore, spectral signatures of $\gamma\gamma$ absorption should be easier to see in a phase-average spectrum if emission occurs close to a black hole.

The absorption spectra expected at various orbital phases, and for the two inclinations considered, are shown in Fig.~\ref{ls_spec}. There is a clear absorption at all orbital phase with limited spectral changes  in the $i=20$\degr\ case (black hole). The shape is roughly independent of phase with maximum absorption occuring around 400~GeV. In contrast, the energy of maximum absorption changes dramatically in the $i=60$\degr\ case, shifting from 0.2-0.3~TeV to 2-3~TeV when going from orbital phase $\phi=0.1$ to $\phi=0.7$. Hence, more dramatic changes are expected in phase-binned spectra if the inclination is high.

The 2004 \hess\ spectrum covers the 0.3-3~TeV range. Weighting by the integrated flux, the {\em phase-averaged} absorption for $i=20$\degr\ would produce a hardening of almost $0.5$ of the intrinsic photon index between 0.3-3~TeV, assuming power-law emission at the source. Of course, a single power-law fit would {\em not} be an adequate description for a high S/N spectrum of the absorbed $\gamma$-rays. When $i=60$\degr, the amplitude of the variations in absorption is greater, but the phase-averaged hardening is slight, $\approx 0.1$ in photon index. Spectral changes are complex in this case, depending upon energy threshold. Above a few TeV, the overall effect is a spectral hardening at low fluxes compared to high fluxes. With a threshold at a few 0.1~TeV, the spectrum may appear to soften at high fluxes as the absorption trough moves to more than a TeV and more low energy photons are detected. For a neutron star, the orbital phase at which the flux peaks provides an almost direct view of the intrinsic spectrum, with at most $\approx$20\% absorption at 2-3~TeV. The intrinsic spectrum is always significantly absorbed ($\tau_{\gamma\gamma}\ga 1$) in the black hole case.

\section{Gamma-ray absorption in \lsi\ and \psrb}

Absorption in \lsi\ and \psrb\ is investigated here under identical assumptions as \ls\ for the gamma-ray emission (constant, isotropic, close to compact object). \psrb\ has been detected by \hess\ near periastron. Inaccessible to \hess\ (but accessible to {\em MAGIC} and {\em VERITAS}), \lsi\ has yet to be detected by ground-based Cherenkov telescopes.

\subsection{\lsi}
\lsi\ is a massive X-ray binary with a very similar spectral energy distribution to \ls. It also has resolved radio emission. \lsi\ has long been associated to hard $\gamma$-ray emission with possible counterparts seen by {\em Cos-B} and {\em EGRET}. This prompted \citet{maraschi} to propose that high energy emission in \lsi\ is powered by an interacting pulsar wind. Jet models have been proposed by \citet{boschparedes} and \citet{romero}. The B0V star has $T_\star\approx 22,500$~K and $R_\star\approx 10$~R$_\odot$. Recently, \citet{casares2} have updated the orbital parameters from optical spectroscopy, finding $P_{\rm orb}\approx 26.496$~days, $e\approx0.72$, $\omega\approx21$\degr. Radial velocity studies are insufficient to rule out a neutron star. As with \ls, two cases are studied, representing extremes compatible with observations; one with $i=60$\degr\ for a 1.4~M$_\odot$ neutron star, and one with $i=20$\degr\ for a 4~M$_\odot$ black hole.

\begin{figure}
\resizebox{\hsize}{!}{\includegraphics{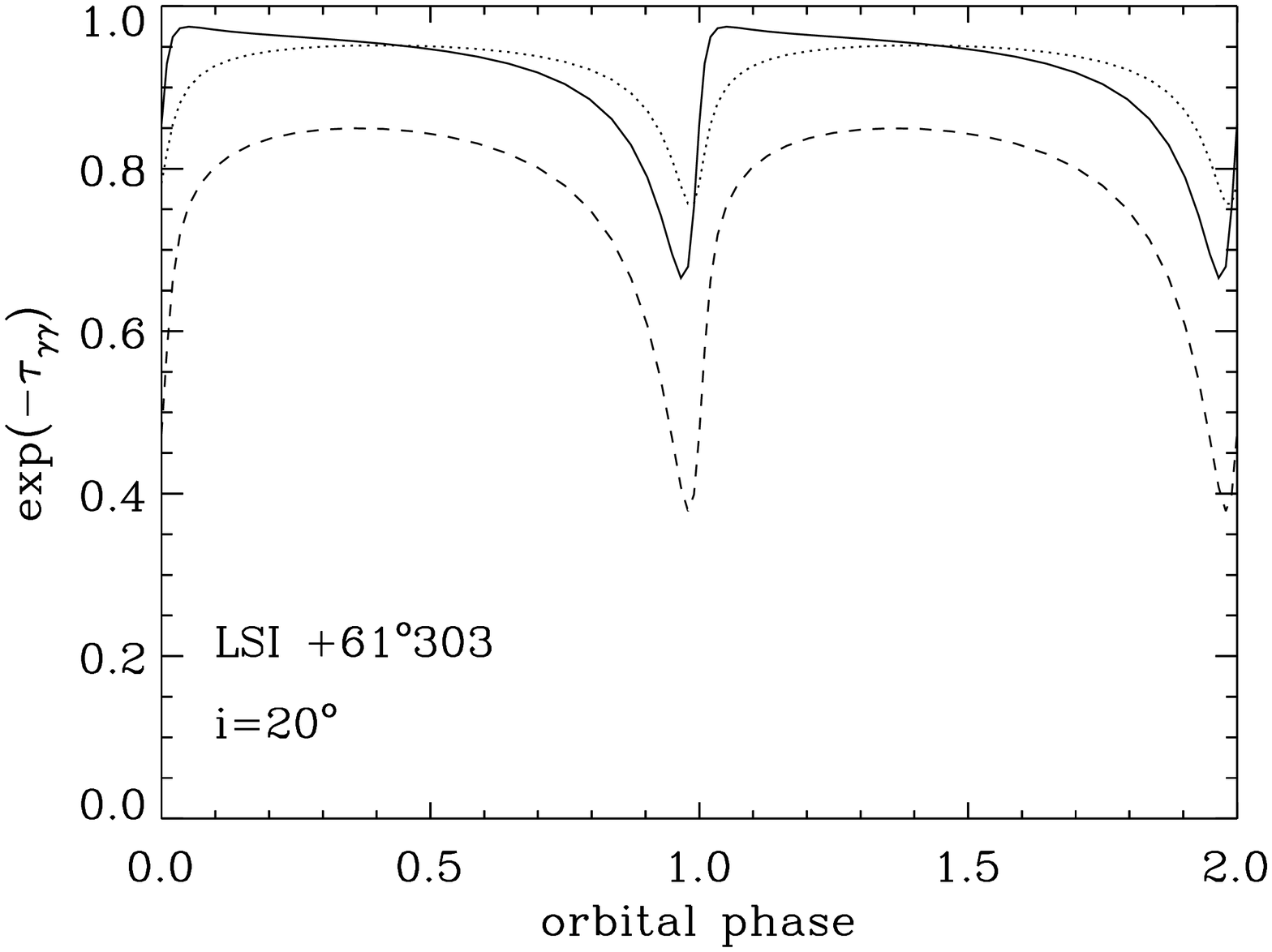}}
\resizebox{\hsize}{!}{\includegraphics{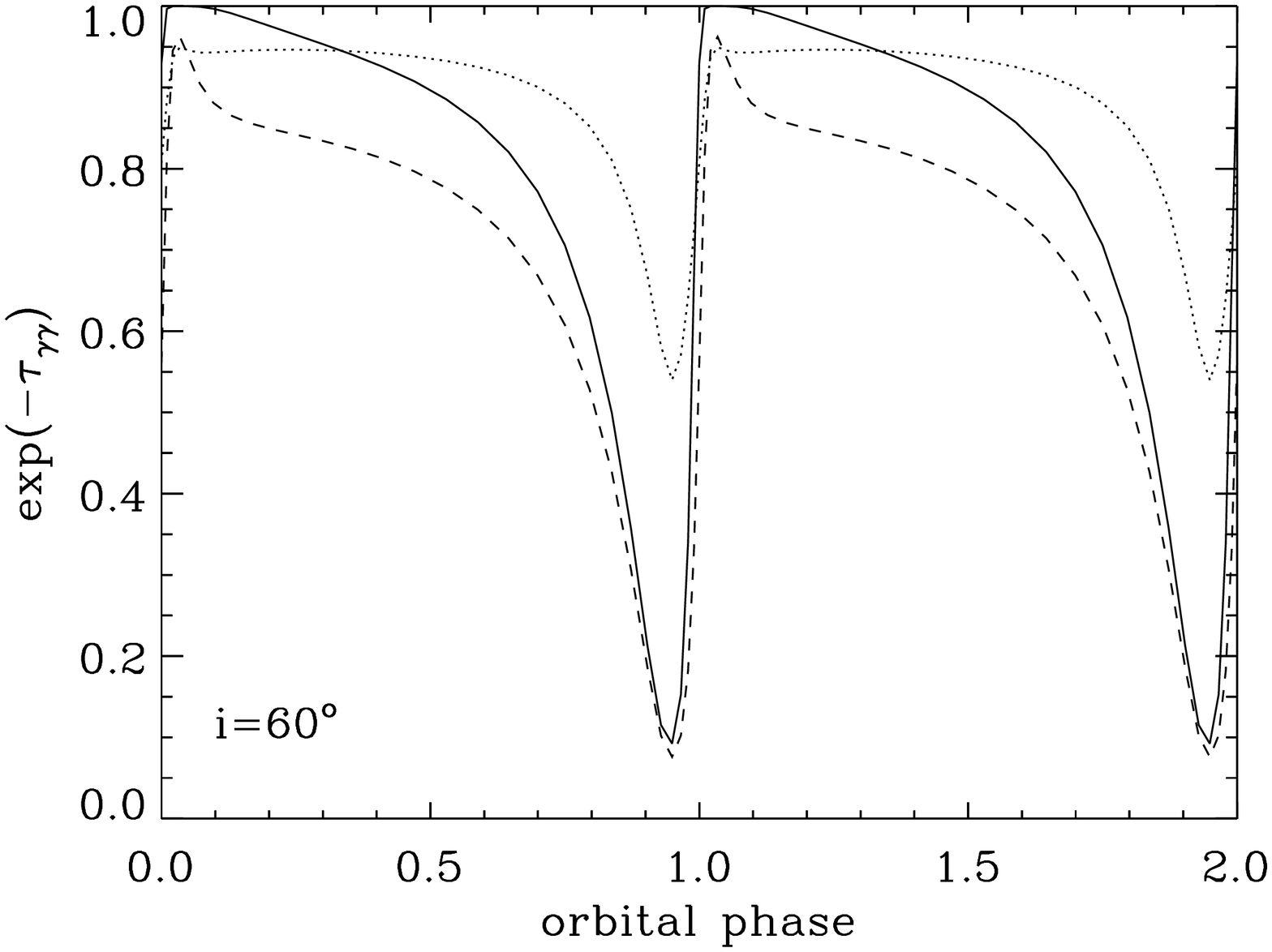}}
\caption{Modulation of the $\gamma$-ray transmission with orbital phase
in \lsi\ for $i=20$\degr\ (4~M$_\odot$ black hole) and $i=60$\degr\ (1.4~M$_\odot$ neutron star). The solid line is for 100~GeV photons, dashed is for 1~TeV and dotted for 10~TeV photons. Significant absorption occurs only slightly before periastron ($\phi=0$) in both cases. As with \ls, a higher inclination leads to a larger amplitude of absorption.}
\label{lsilc}
\end{figure}

The absorption lightcurve is shown for both cases in Fig.~\ref{lsilc}. The orbit is wider than in \ls\ so that the variation in $\gamma$-rays absorption occurs essentially only close to periastron. The distance to the star is then $\approx 0.1$~AU, as with \ls. The stellar luminosity is lower in \lsi\ than in \ls, so that not all the $\gamma$-rays are absorbed at periastron despite the similar distance. At apastron the orbital separation is $\approx 0.7$~AU but there is still some absorption, around $20\%$ of the 1~TeV flux, because the compact object is behind the star when projecting the orbit onto the sky.

The amplitude of the variations is greater at high inclinations, as in \ls. The spectral changes expected close to periastron for $i=60$\degr\ are dramatic, with peak emission that follows the minimum within a few days. Fig.~\ref{lsilc} shows the strong hardening of the spectrum above 1~TeV with lower fluxes. The absorption feature outside of periastron is a 20\% effect. \lsi\ suffers less absorption for most of its orbit so should be more easily detected at VHE energies than \ls, if the same processes are indeed at work in the two.

\begin{figure}
\resizebox{\hsize}{!}{\includegraphics{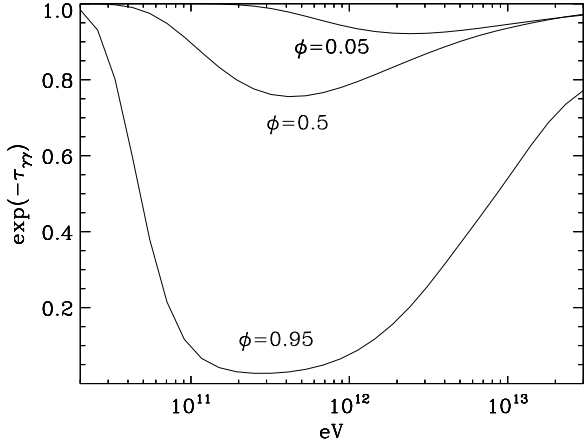}}
\caption{\lsi\ absorption spectrum at orbital phases $\phi=0.05$,0.5 and 0.95 assuming $i=60$\degr\ (see bottom panel of Fig.~\ref{lsilc}). There is a dramatic spectral change within a couple of days of periastron.}
\end{figure}

\subsection{\psrb}
\begin{figure}
\resizebox{\hsize}{!}{\includegraphics{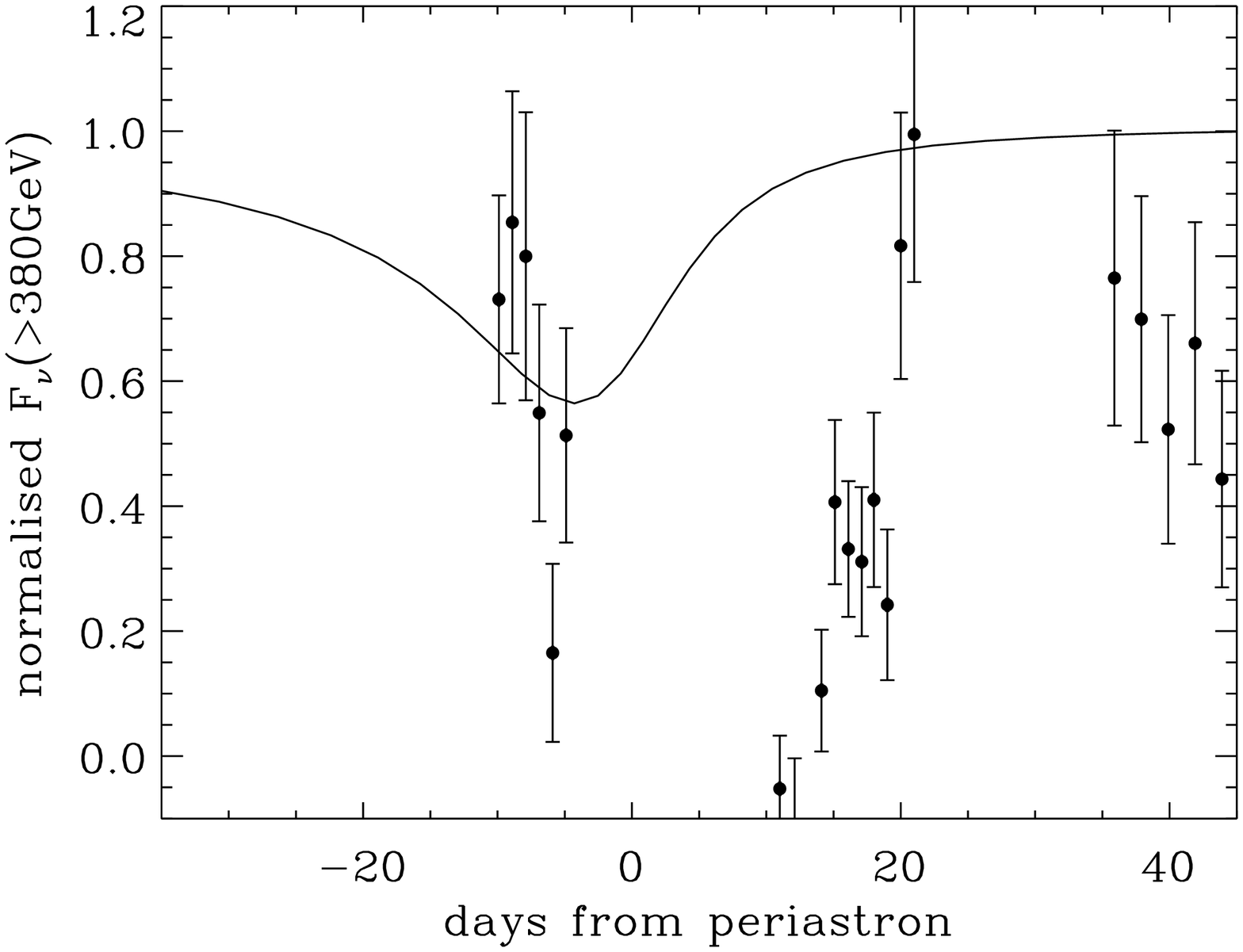}}
\caption{Modulation caused by $\gamma$-ray absorption around periastron in \psrb. The data points are the \hess\ integrated lightcurve above 380~GeV measured during the 2004 periastron passage \citep{psrb}. The data are renormalised to the maximum $10^{-7}$~m$^{-2}$~s$^{-1}$. The solid line is the integrated flux above the same threshold after absorption, assuming constant power law emission wih a photon index of -2.8. Absorption plays a minor role in the observed variability.}
\label{psrblc}
\end{figure}
Pulsar timing in \psrb\ allows an accurate determination of orbital parameters: $P_{\rm orb}\approx 1236.7$~days, $e\approx0.87$, (star) $\omega\approx138.7$\degr, $i\approx 35$\degr. The star has $T_\star\approx 27,000$~K and $R_\star\approx 10$~R$_\odot$ for a total luminosity of 5.8~$10^4$~L$_\odot$ \citep{johnston}. Fig.~\ref{psrblc} shows the expected lightcurve around periastron passage, taking only absorption into account. The $\gamma$-rays are emitted at the pulsar location with a power law spectrum $dN_\gamma\propto E^{-2.8}dE$. The pulsar is $\approx 10^{13}$~cm at its closest to the Be star, further away than \ls\ or \lsi. Although the pulsar is $\ga$20~$R_\star$ away from the star, there is still some absorption because of the orbital orientation; photons emitted at $\phi\approx 0$ must travel near the star on their way to the observer.

Maximum absorption occurs slightly before periastron, with up to 40\% of the $\gamma$-rays above the \hess\ threshold (380~GeV) being absorbed. The measured \hess\ lightcurve has a greater amplitude with a minimum occuring later than expected. Absorption is clearly not sufficient to explain the observations and some of the variability must be intrinsic. A likely possibility is the angle-dependence in the inverse Compton interactions producing the TeV emission \citep{kirk}. VHE emission would not be isotropic. Models should be compared to absorption corrected fluxes. Finally, hardening by absorption of the 0.3-3~TeV spectrum is at most 0.2 in photon index, too small to be measured by \hess.

\section{\ls\ and \lsi: opacity in a jet ?}
Are there good reasons to expect that $\gamma$-rays are emitted at the location of the compact object? For a binary plerion like \psrb, most of the VHE emission should occur close to the pulsar wind termination shock \citep{tavani,kirk}. The overall geometry is set by the stagnation point $R_s$, the point on the orbital axis where ram pressure from the stellar and pulsar wind balance \citep{maraschi}. For \ls, $(R_s/d) \approx 0.1\dot{E}_{36}^{1/2}$ where $d$ is the orbital separation (0.1-0.2~AU in \ls) and  $\dot{E}_{36}=10^{36}$~erg~s$^{-1}$ is the pulsar spindown power (taken equal to that measured in \psrb). The assumed stellar wind has a mass loss rate of $10^{-7}$~M$_\odot$~yr$^{-1}$ and a coasting speed $v_w=2000$~km~s$^{-1}$, appropriate to \ls\ (C05). The stagnation point in \psrb\ is also at about a tenth of the orbital separation \citep{tavani}. VHE emission in this case can be expected to occur on a small scale compared to the orbital separation, close to the pulsar.

There is no such certainty with a black hole (or neutron star) jet. Jet models for the $\gamma$-ray emission have been proposed for both \ls\ and \lsi\ \citep{boschparedes,romero,pjet}. Although the gravitational potential energy reservoir that can be tapped is largest close to the black hole, most of the energy may not be radiated there if it is carried by protons or Poynting flux \citep{sikora}. Shocks or reconnection can convert kinetic or electromagnetic energy into radiation progressively over large distances, up to a termination shock with the ISM. There is evidence for X-ray synchrotron emission on parsec scales from high energy electrons in the relativistic jet of \xte\ \citep{corbel}.

\subsection{Variation with height}
\begin{figure}
\resizebox{\hsize}{!}{\includegraphics{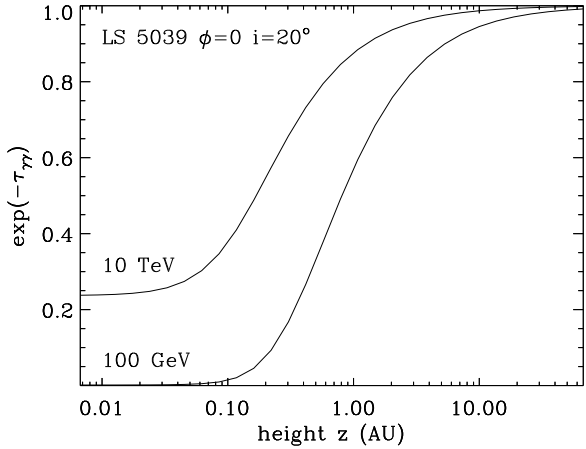}}
\caption{Variation in transmitted flux with height $z$ above the plane in \ls. The assumed jet is perpendicular to the orbital plane, the inclination is 20\degr\ and the black hole is at periastron. The absorption is shown for 100~GeV and 10~TeV photons. The curve for 1~TeV photons would be indistinguishable from the 100~GeV curve in this plot. VHE photons emitted a few AU away from the black hole suffer little absorption. For comparison purposes, the orbital separation in \ls\ varies between 0.1 and 0.2~AU.}
\label{jet}
\end{figure}

The change in flux absorption with height above the orbital plane in \ls\ is shown in Fig.~\ref{jet}. The inclination is 20\degr\ and the black hole is at periastron. Absorption decreases to less than $50$\% at all energies when emission occurs more than $\approx 0.7$ AU away from the plane, as discussed in \citet{aharonian,science} and \citet{dermer}. The same calculation for \lsi\ yields a distance of $\approx 0.1$~AU.  If VHE emission can be independently linked to a relativistic jet, a lack of temporal or spectral absorption signatures will imply emission at large distances. 

Unlike TeV blazars, the target photon field here is external to the jet and $\gamma\gamma$ absorption will not be reduced by relativistic aberration. At a given energy, the dependence of the opacity with height is set by the companion light, providing constraints on models as in \citet{blandford} where the jet is structured by the $\gamma$-photosphere. 

\subsection{Beaming}
The height at which $\tau_{\gamma\gamma}<1$ in \ls\ corresponds to roughly $10^7$ gravitational radii $R_g$ for a stellar mass black hole. In blazars, VHE emission is often proposed to occur in the broad line region at parsec-scales, {\em i.e.} about $10^4~R_g$ for a $10^9$~M$_\odot$ black hole. In the internal shock model, the distance of first interaction is $\sim\Gamma^2 c\Delta t$ where $\Delta t$ is the observed timescale and $\Gamma$ is the typical bulk Lorentz factor of ejecta \citep{spada}. Assuming \ls\ is a genuine microblazar with scaled-down physics, then $\Delta t$ would be reduced with black hole mass. For the given emission distances, the conclusion would be that $\Gamma$ in \ls\ is at least the same, if not higher than in blazars.

Could ejections in \ls\ have $\Gamma\approx 10$, as inferred in blazars from opacity arguments and $\gamma$-ray modelling \citep{dondi,ghisellini}?  Emission from a jet will be {\em de}boosted if viewed at an angle larger than $\approx 25$\degr\ for $\Gamma$=10. The system inclination required for a black hole primary is close to this value. In this case, emission from a jet perpendicular to the orbital plane would not be boosted. This might fit with the weak X-ray to $\gamma$-ray luminosity observed ($\sim 10^{34}$~erg~s$^{-1}$). Then, \ls\ would not be a genuine {\em microblazar} with the jet along the line-of-sight. 

Current VLBI radio observations of \ls\ only indicate a mildly relativistic outflow with $v\la0.4c$ \citep{paredes}. Yet, the Lorentz factors of TeV blazars deduced from $\gamma\gamma$ absorption are known to be much greater than those measured at larger distances by VLBI observations \citep{piner}. In both cases, there would be a puzzling discrepancy between the high $\Gamma$ inferred from $\gamma$-ray studies and the low $\Gamma$ measured on larger scales in radio.

\section{Effect of pair production}
The $e^+e^-$ pairs created by absorption can boost stellar photons to very high energies via inverse Compton interactions. These $\gamma$-rays can in turn be absorbed and a cascade follows. The initial pair created by absorption of a 1~TeV photon has a Lorentz factor $\gamma \approx 10^6$. In \ls, their inverse Compton interaction with stellar photons occurs in the Klein-Nishina regime on a timescale $t_{\rm KN}\approx8\gamma_6 d_{0.1}^2 \left[\ln \gamma_6 +1.2 \right]~\mathrm{seconds}$ \citep{bg}, with $d_{0.1}$=0.1~AU. Hence, the typical distance travelled by the pair before a photon is boosted to high energies is small enough compared to the binary size to start a cascade. However, the cascade will be quenched if the pair loses energy preferentially to another emission process that does not re-emit VHE $\gamma$-rays (e.g. synchrotron; \citealt{bednarek97,aharonian}).

The cascade reduces the effective $\tau_{\gamma\gamma}$, since absorbed $\gamma$-rays are partly re-radiated (at lower energies). If $\tau_{\gamma\gamma}$ is very high, energy initially radiated in the absorbed band is degraded to frequencies below the pair production threshold ($\la 30$~GeV in \ls). The pile up at these frequencies might account for the luminosity step between the {\em EGRET} (GeV energies) and {\em HESS} (TeV energies) measurements (see Fig.~1 in \citealt{aharonian}). Absorption of a constant spectrum in $\nu F_\nu$ from GeV to TeV energies cannot account for this difference. Such a spectrum would, on average, only be absorbed by a factor 3-10 above 250~GeV (\S3.1) when the luminosity jump is $\approx$30. Taking into account the photons degraded to lower energies by the cascade would raise the absorption step. Orbital variability would then even be expected at energies below the interaction threshold ($\la 30$~GeV, hence in the {\em EGRET} or {\em GLAST} range), in antiphase with the absorption lightcurves. \citet{bednarek97} called this ``$\gamma$-ray focusing by soft radiation''. 

Angular effects must be considered for $\gamma\gamma$ opacities of a few. The estimate of the Klein-Nishina interaction timescale assumes an isotropic photon bath. Just as the $\gamma\gamma$ absorption is changed by the geometry, the pair motion with respect to the star will change the inverse Compton opacity. Maximum boost will be achieved for head-on collisions with stellar photons. Because of momentum conservation, the cascade should tend to develop along the direction where the initial photon was emitted. However, the cascade is effectively isotropized as the gyroradius is small compared to the interaction length \citep{bednarek97}. Radiation can then be redistributed at other orbital phases.

A quantitative description of these effects is feasible by Monte-Carlo methods \citep{bednarek97,bednarek00,bednarek} but is beyond the scope of this work. A model of the intrinsic spectrum should be considered before doing this. Cascades will probably be important only when the $\gamma\gamma$ opacity is high and the source behind the star in the plane of the sky. This ensures maximum energy boost and maximum density of stellar photons along the initial development of the cascade. This is the case at periastron in \ls\ but not in \lsi\ (source close to front of the star at maximum absorption) and even less in \psrb\ (low $\tau_{\gamma\gamma}$). The amplitude of  flux variations will be smaller, though perhaps not noticeably so on a linear scale and with energy thresholds $\ga 100$~GeV, which are usually at or above the energy of maximum absorption (e.g. compare the Cen X-3 lightcurves above 300~GeV with and without cascade in \citealt{bednarek00}). The absorption spectrum around periastron will be softened. The changes will be small when $\tau_{\gamma\gamma}\la 1$. Qualitative differences in absorption at TeV energies for a neutron star or black hole case should therefore remain.

\section{Conclusion}
The intense stellar radiation field in massive X-ray binaries provides a novel way of investigating very high energy processes around compact objects. Because of the dependence of the $\gamma\gamma$ cross-section on angle, the anisotropy of the field has a strong effect on the opacity to very high energy $\gamma$-rays $\ga10$~GeV. The orbital orientation to the observer and location of emission are crucial in determining the absorption. In \psrb\, this leads to significant absorption before periastron, although the pulsar is $>20$R$_\star$ from its stellar companion.

The changing radiation field modulates the gamma-rays in a predictable fashion. The result is a controlled, repeatable experiment that can help pinpoint the location of emission. Comparing the predictions to observations is essential to single out the variability at the source from that due to absorption in the binary. In \ls, the binary separation is only 2-4~$R_\star$ yet absorption can be much smaller than implied by simple estimates. Indeed, absorption of $\gamma$-rays emitted close to a neutron star (implying a high inclination) is negligible at inferior conjunction. The spectrum above 1~TeV is considerably hardened near periastron in \ls\ and \lsi. \lsi\ is largely absorption-free which should ease its detection at high energies by {\em MAGIC} or {\em VERITAS}, if similar processes are at work in both sources. 

The detection of an orbital modulation of the TeV flux in \ls\ would imply emission within the system, either from a pulsar wind termination shock or from the inner region of a jet. If the primary is a black hole, the inclination will be $\approx 20$\degr\ and the absorption spectrum has a break at $\approx$~400~GeV with limited orbital spectral changes. For a neutron star, the inclination is $\approx 60$\degr\ and the break is expected to change from 0.2 to 2~TeV from low to high fluxes. The absence of orbital variations or spectral signatures in VHE observations of \ls\ would suggest emission occurring in regions of low $\gamma\gamma$ opacity. Jet emission at large distances compared to the system scale would be the prime candidate.

\begin{acknowledgements}
I thank B. Giebel, P. Bruel, J.-P. Lasota, V. Bosch-Ramon and M. Ribo for their helpful comments on this work and the organisers of the KITP "Physics of Astrophysical Outflows and Accretion Disks" program where part of this work was completed. This research was supported in part by NSF grant No. PHY99-07949.
\end{acknowledgements}

\appendix

\section{Calculation of $\gamma$-ray absorption in a massive X-ray binary system}

\begin{figure}
\resizebox{7cm}{!}{\includegraphics{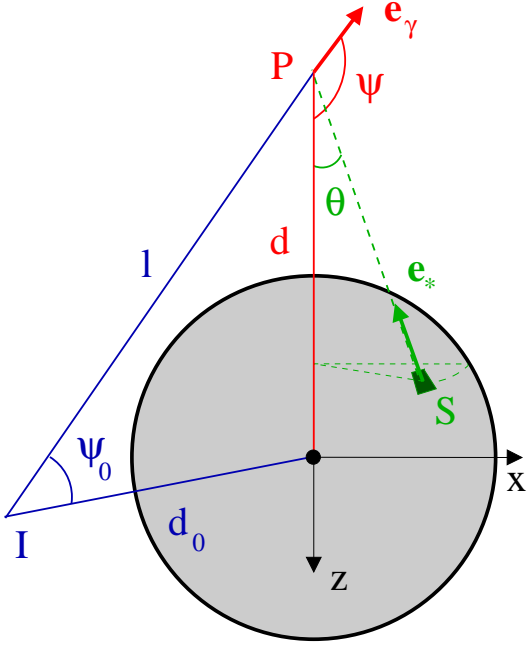}} 
\caption{Geometry for absorption of a $\gamma$-ray at a location $P$ due to $e^+e^-$ pair creation with stellar photons emitted at $S$. The $\gamma$-ray is emitted at $I$ and $l$ is the length of the $\gamma$-ray path to $P$. The ($x$,$z$) plane is defined by the star center and $\gamma$-ray path.}
\label{geometry}
\end{figure}

\subsection{Opacity in vicinity of the star}
The differential absorption opacity seen by a $\gamma$-ray of energy $E$ located at a position P and travelling in the direction $\mathbf{e}_\gamma$ due to photons of energy $\epsilon$ emitted at S along $\mathbf{e}_\star$ is \citep{gs}
\begin{equation}
\mathrm{d}\tau_{\gamma \gamma}=  (1-\mathbf{e}_\gamma \mathbf{e}_\star) n_\epsilon  \sigma_{\gamma \gamma} \mathrm{d}\epsilon \mathrm{d}\Omega \mathrm{d}l
\end{equation}
$\mathbf{e}_\gamma$ and $\mathbf{e}_\star$ are unit vectors in the directions of propagation;  $\mathrm{d}\Omega$ is the solid angle of the surface emitting the $\mathbf{e}_\star$ photons; $n_\epsilon$ is the radiation density in
photons~cm$^{-3}$~erg$^{-1}$~sr$^{-1}$.
%The photon density along the $\gamma$-ray path $\mathrm{d}l$ is $(1-\mathbf{e}_\gamma \mathbf{e}_\star) n_\epsilon \mathrm{d}\epsilon \mathrm{d}\Omega$.

The $\gamma\gamma$ interaction cross-section $\sigma_{\gamma \gamma}$ depends only on $s$, defined as \citep{gs}
\begin{equation}
s=\frac{\epsilon E}{2 m_e^2 c^4}(1-\mathbf{e}_\gamma \mathbf{e}_\star)
\end{equation}
and the photons interact only if $s>1$. The variable $s$ can be rewritten as $s$=$\epsilon/\epsilon_{\rm min}$ where $\epsilon_{\rm min}$ is the threshold for interaction at $\gamma$-ray energy $E$ (Eq.~\ref{eq:threshold}). 

The massive star dominates the radiative output in \ls, \lsi\ and \psrb\ so that any other source of radiation for pair-production with TeV $\gamma$-rays is neglected. The star has a radius $R_\star$ and is assumed to have a blackbody radiation density with a temperature $T_\star$
\begin{equation}
n_\epsilon = \frac{2 \epsilon^2}{h^3 c^3} \frac{1}{\exp (\epsilon/kT_\star) -1} ~~\mathrm{ph~cm}^{-3} \mathrm{erg}^{-1} \mathrm{sr}^{-1}
\end{equation}

Orienting the coordinate system so that the $\gamma$-ray path is in the ($x$,$z$) plane (Fig.~\ref{geometry}), then the location and direction of the $\gamma$-ray are given by the angle $\psi$ and distance $d$ to the star center and
%\begin{equation}
%\mathbf{e}_\gamma \cdot \mathbf{e}_\star=\left( \begin{array}{c} \sin \psi \\ 0 \\ \cos\psi \end{array} \right) \cdot \left(\begin{array}{c} -\cos\phi \sin \theta \\ -\sin\phi \sin \theta \\ -\cos\theta \end{array} \right)
%\end{equation}
\begin{equation}
1-\mathbf{e}_\gamma \cdot \mathbf{e}_\star= 1+\cos \psi \cos \theta +\sin\psi \cos \phi \sin \theta
\end{equation}
The solid angle of the star surface element emitting photons along $\mathbf{e}_\star$ is
\begin{equation}
\mathrm{d}\Omega=\sin \theta \mathrm{d}\phi \mathrm{d}\theta
\end{equation}

The length $l$ of the $\gamma$-ray path since its emission at a distance $d_0$ and at an angle $\psi_0$ (point I in Fig.~\ref{geometry} with $\psi_0$=0 for a $\gamma$-ray travelling directly towards the center) is related to $\psi$ so that:
%\begin{eqnarray}
%\psi &=& \tan^{-1} \left(\frac{d_0 \sin \psi_0}{d_0\cos \psi_0 - l}\right) ~~~(l<d_0\cos \psi_0)\\
%\psi &=& \pi+ \tan^{-1} \left(\frac{d_0 \sin \psi_0}{d_0\cos \psi_0 - l}\right) ~~~(l>d_0\cos \psi_0)
%\end{eqnarray}
\begin{equation}
\psi = \tan^{-1} \left(\frac{d_0 \sin \psi_0}{d_0\cos \psi_0 - l}\right) ~~~\mathrm{for~} l<d_0\cos \psi_0
\end{equation}
and $\psi=\pi+\tan^{-1}(...)$ for $l>d_0\cos \psi_0$. The distance $d$ is related to $l$ by
\begin{equation}
d^2=d_0^2+l^2-2d_0 l \cos\psi_0
\end{equation}

The dependence of the differencial opacity
$\mathrm{d}\tau_{\gamma\gamma}$ on $E$, $\epsilon$, $\phi$, $\theta$
and $l$ is now explicit. The total opacity to infinity for a
$\gamma$-ray of given energy $E$ and initial location requires a
quadruple integral:
\begin{equation}
\tau_{\gamma\gamma}=\int^{\infty}_0 \mathrm{d}l \int^{1}_{c_\mathrm{min}} \mathrm{d}\cos \theta \int^{2\pi}_{0} \mathrm{d}\phi \int^{\infty}_{\epsilon_\mathrm{min}} \frac{\mathrm{d}\tau_{\gamma\gamma}}{\mathrm{d}\epsilon\mathrm{d}\Omega\mathrm{d}l} \mathrm{d}\epsilon 
\label{full}
\end{equation}
where $c_{\rm min}=(1-R^2_\star/d^2)^{1/2}$. It is convenient to rewrite the energy integral as a definite integral on $\beta=(1-1/s)^{1/2}$ within the interval $[0,1]$. Similarly, the integral on $l$ can be transformed into a definite integral on $\psi$ between $[\psi_0,\pi]$ with the condition $d_0 \sin \psi_0\ge R_\star$ (or the $\gamma$-ray hits the star). An iterated Gaussian quadrature is used to calculate the integral.

\begin{figure}
\resizebox{\hsize}{!}{\includegraphics{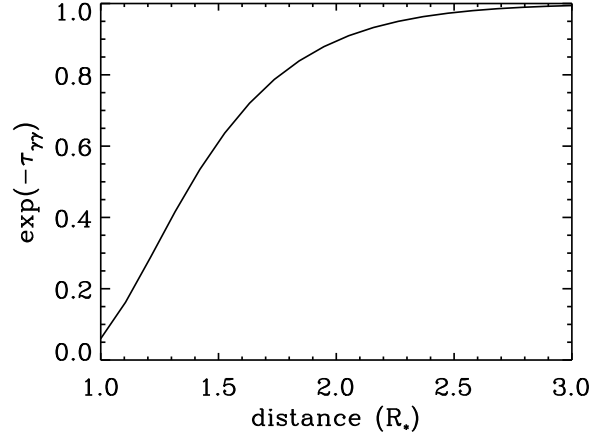}} 
\caption{For a $\gamma$-ray emitted directly away from the star ($\psi=\pi$), the point source approximation (Eq.~\ref{approx}) predicts no absorption. The exact transmitted flux $\exp(-\tau_{\gamma\gamma})$, calculated using Eq.~\ref{full} is shown here for a 1~TeV $\gamma$-ray as a function of the distance of its emission point ($R_\star$=10~R$_\odot$, $T_\star$=40,000~K). Because of the star's finite size, there is absorption even for $\psi=\pi$ when the $\gamma$-ray is at distances $\la 2R_\star$.}
\label{num}
\end{figure}

When $R_\star\ll d$, an integration of Eq.~\ref{full} over $\theta$ and $\phi$ returns the point source approximation
\begin{equation}
\mathrm{d}\tau_{\gamma\gamma}=\pi\left(\frac{R_\star}{d}\right)^2 \sigma_{\gamma\gamma} n_\epsilon (1+\cos\psi) \mathrm{d}l\mathrm{d}\epsilon 
\label{approx}
\end{equation}
The threshold is $\epsilon_{\rm min}=2m_e^2c^4/E(1+\cos\psi)$. Numerical comparisons show the approximation works reasonably well except for $\psi\approx \pi$ (Fig.~\ref{num}) or when $d-R_\star<R_\star$ at closest approach. The full calculation is then preferable. The results presented here were derived using the exact calculation.

\subsection{Orbital dependence}
Assuming emission close to the compact object, the initial distance $d_0$ of the $\gamma$-ray to the star is simply
\begin{equation}
d_0=\frac{a(1-e^2)}{1+e\cos(\theta-\omega)}
\end{equation}
with the semi-major axis $a=(G M P^2_{\rm orb}/4\pi^2)^{1/3}$ ($M$ is the total mass), $e$ the eccentricity, $\theta$ the true anomaly and $\omega$ the periastron angle.  The orbital phase (mean anomaly) is $\phi=(\eta-e \sin\eta)/2\pi$ where $\eta$ is the eccentric anomaly with
\begin{equation}
\tan \frac{\eta}{2}=\left(\frac{1-e}{1+e}\right)^{1/2} \tan\frac{\theta-\omega}{2}
\end{equation}
The angle $\psi_0$  of emission to the observer (located at $\infty$) measured from the star, at each point of the orbit (Fig.~\ref{geometry}) is
\begin{equation}
\cos \psi_0={\sin \theta}{\sin i}
\end{equation}
where $i$ is the inclination of the orbit. If emission occurs at a height $z$ above the location of the compact object and perpendicular to the orbital plane (as in Fig.~\ref{jet}), then the initial distance to the star simply becomes $d=(d_0^2+z^2)^{1/2}$ and the initial angle is changed to $\cos \psi = (d_0/d) \cos \psi_0$ (with $d_0$ and $\psi_0$ given above).
\end{document}